\documentclass[reprint, floatfix,
 amsmath,amssymb,
 aps,
 prapplied,
]{revtex4-2}
\usepackage{textgreek}
\usepackage{graphicx}
\usepackage{dcolumn}
\usepackage{bm}

\begin{document}

\preprint{APS/123-QED}

\title{Graphene-Enhanced Single Ion Detectors for Deterministic Near-Surface Dopant Implantation in Diamond}
\author{Nicholas F. L. Collins}
 \email{ncollins2@student.unimelb.edu.au}

\author{Alexander M. Jakob}%
\author{Simon G. Robson}%
\author{Shao Qi Lim}%
\author{Jeffery C. McCallum}%
\author{David N. Jamieson}%

\affiliation{%
 Centre for Quantum Computation and Communication Technology, School of Physics, University of Melbourne,
Parkville, Victoria 3010, Australia
}%


\author{Boqing Liu}%
\author{Yuerui Lu}
\affiliation{School of Engineering, College of Science and Computer Science, ANU }%

\author{Paul R\"acke}%
\author{Daniel Spemann}%
\affiliation{%
 Leibniz Institute of Surface Engineering (IOM), Permoserstr. 15, Leipzig D-04318, Germany
}%

\author{Brett C. Johnson}%
\affiliation{%
 School of Science, RMIT University
}%

\date{\today}

\begin{abstract}
\noindent Colour centre ensembles in diamond have been the subject of intensive investigation for many applications including single photon sources for quantum communication, quantum computation with optical inputs and outputs, and magnetic field sensing down to the nanoscale. Some of these applications are realised with a single centre or randomly distributed ensembles in chips, but the most demanding application for a large-scale quantum computer will require ordered arrays. By configuring an electronic-grade diamond substrate with a biased surface graphene electrode connected to charge-sensitive electronics, it is possible to demonstrate deterministic single ion implantation for ions stopping between 30 and 130~nm deep from a typical stochastic ion source. An implantation event is signalled by a charge pulse induced by the drift of electron-hole pairs from the ion implantation.  The ion implantation site is localised with an AFM nanostencil or a focused ion beam. This allows the construction of ordered arrays of single atoms with associated colour centres that paves the way for the fabrication of deterministic colour center networks in a monolithic device.
\end{abstract}

\maketitle


\section*{\label{sec:level1}Introduction}

\begin{figure}[t]
    \centering
    \includegraphics[page=1,width=
    0.97\linewidth,trim={0cm 0cmcm 0cm 0}]{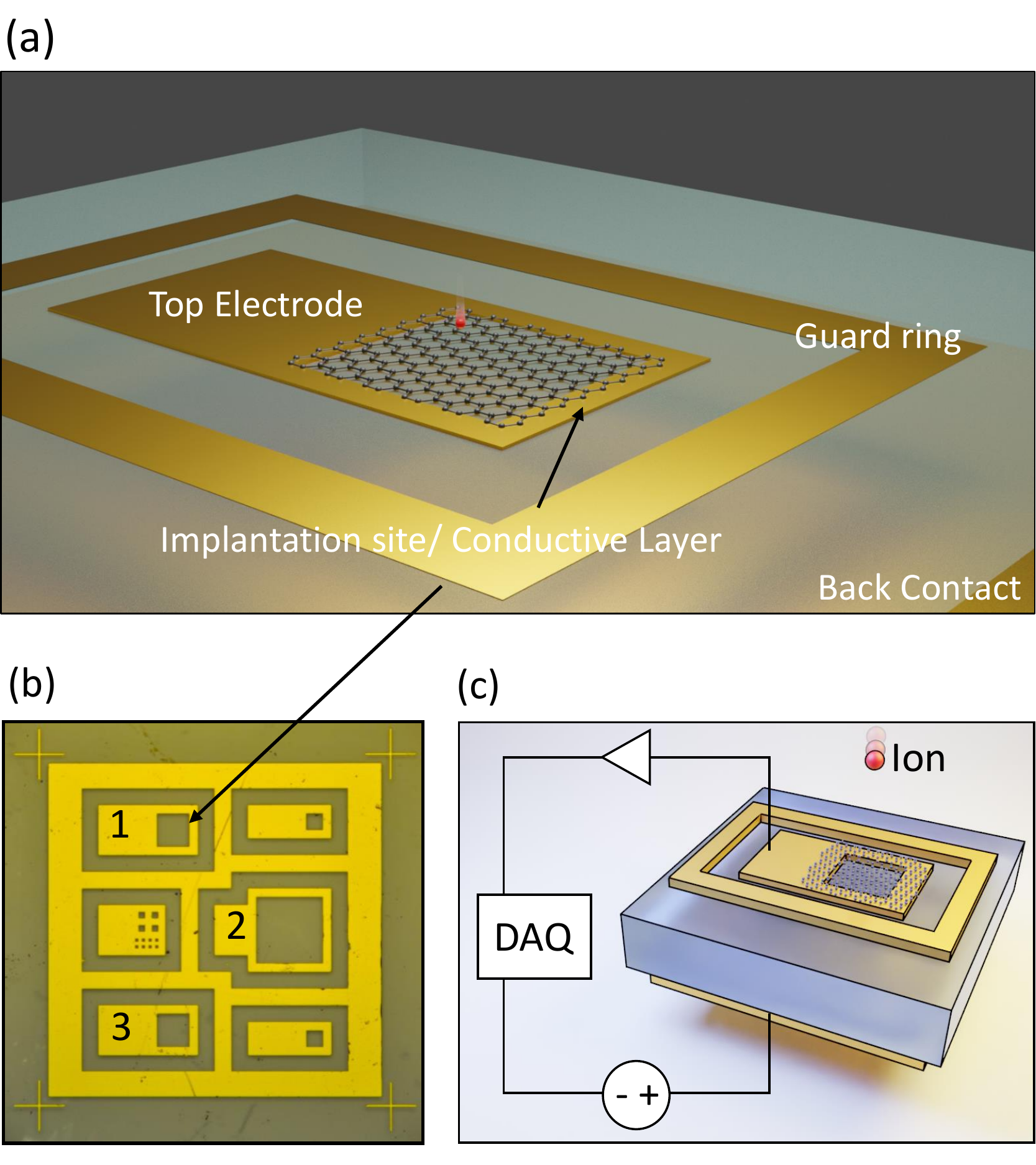}
    \caption{(a) Schematic of a single ion detector. The metal top electrode surrounds a rectangular implantation site in which atoms can be deterministically implanted. The site is coated with a graphene layer (illustrated via a stick and ball model). (b) Optical top view on a diamond substrate featuring multiple single ion detectors with a common outer guard ring that suppresses surface leakage into the detector regions. Numbers correspond to electrodes used for testing. (c) Cross-sectional schematic of the sandwich detector electrode design with a top and back metal electrode on respective substrate surface. A thin conductive layer (black line) covers the surface of the implantation site embedded in the top electrode. The top electrode is connected to an ultra-low-noise charge-sensitive pre-amplifier with subsequent pulse processing electronics. The back electrode is biased to a high electric potential, forming a strong drift field for ion-induced charge carriers inside the diamond substrate. 
    }
    \label{fig:Figure1}
\end{figure}

The scalable fabrication of diamond-based quantum technologies requires the controlled placement of individual defects within 30 nm laterally for entangled colour centres, and between 50 and 100 nm when placed in appropriate cavities \cite{Dolde2013,Smith2019}. 
For quantum memories, colour centers have to be  embedded in optical cavities or resonators, whereas quantum processors demand close proximity between centres \cite{Bradley2019,Rozpedek2019,Neumann2010}. For these devices, colour centers that have promising attributes include the nitrogen vacancy (NV), the silicon vacancy (SiV), the germanium vacancy (GeV) and tin the vacancy (SnV)  \cite{Doherty2013,Neu2011,Iwasaki2015,Iwasaki2017,Pezzagna2021}.
In the present work we do not address the process for the conversion of an implanted ion into a colour centre which is still a work in progress.  \\
The deterministic doping of diamond substrates with ion implantation requires control over the ion placement at specific locations as well as the number of ions implanted. The controlled placement of implanted ions can be achieved with a Focused Ion Beam (FIB) or with a scanned mechanical nanostencil \cite{Tamura2014,Schroder2017,Meijer2008}. Several technologies are being investigated for precise control of the number of implanted ions \cite{Groot-Berning2019, Racke2022,Cassidy2021,Jakob2022}. 
For silicon substrates, high confidence (99.85\%) has been achieved with single ion detectors integrated on chip\cite{Jakob2022}. Sub-20~nm depth $^{31}$P implantation has been demonstrated. In the present study, we employ a similar approach adapted to diamond substrates.
\\
To apply this method to diamond, metal electrodes were deposited on the top and bottom  of the substrate (sandwich design). These were electrically biased to form a drift field across the substrate for operation analogous to a solid state radiation detector (see Fig. \ref{fig:Figure1}). Upon implantation of a single ion, its kinetic energy is dissipated via nuclear and electronic stopping processes in the diamond lattice. The electronic stopping fraction induces free electron-hole pairs, which in turn drift towards the surface electrodes in the applied electric field. The current transient induced on the electrode by charge drift is collected by a charge sensitive pre-amplifier to produce a signal pulse with an amplitude propositional to the integrated charge induced on the electrode. This technique, known as Ion Beam Induced Charge (IBIC), has the advantage of providing a pulse height spectrum that is a function of the ion mass and kinetic energy. Detailed analysis of individual charge pulse signals can yield information about the ion placement depth \cite{Jakob2022}. This allows for real-time fidelity analysis of implanted atom(s) arrays and the selection of specific atom implants to fulfill placement constraints of certain control schemes, which may increase the yield of colour center networks \cite{Luhmann2019}.\\ 
The IBIC technique was originally developed for mapping the charge collection efficiency in electronic devices employing scanned focused beams of MeV ions \cite{Vittone2004,Vittone2008}. Adapting the technique to sub-30~keV ions required for shallow (sub-20~nm) ion  implantation is challenging because of the far fewer electron-hole pairs induced by each ion impact, particularly considering diamond substrates with a much higher band gap energy than silicon. The ion induced signal is directly proportional to the ion kinetic energy and consequently keV ions produce signals weaker by about two orders of magnitude compared to MeV ions. To meet the challenge requires an ultra-low device leakage current ($\approx$1~pA) and capacitance ($\approx$ 0.5~pF) to retain an acceptable signal-to-noise ratio. For diamond, the use of ultra-pure electronic grade substrates is essential for suppressing bulk charge recombination and preserving the ion induced signal from the on-chip surface electrodes. In this context, the charge collection efficiency describes the fraction of ion-induced free charge carriers in diamond that is detected. State-of-the-art electronic grade diamond substrates have less than 1 ppb N and B content that facilitates long bulk free carrier lifetimes ($1\times10^{-6}$ s). However, as will be shown in this study, further detector enhancements are necessary to address the dominant charge recombination mechanism which arises from surface/interface charge traps.\\
For silicon-based devices, surface charge traps are passivated with a thin, thermally grown, high-quality surface oxide layer \cite{Jakob2022,Robson2022}. However, this method is not implementable with diamond because of the lack of a suitable passivation layer due to the gaseous nature of room temperature CO$_2$. Exploiting insulator thin films based on Al$_2$O$_3$ or SiO$_2$ passivation layers have not been demonstrated to achieve a sufficient interface defect density reduction. To date, oxide interfaces on diamond yield defect concentrations that result in very low carrier mobility \cite{Masante2021}, whilst bare oxygen terminated diamond contains many functional groups and therefore the same behaviour \cite{Ristein2006,Krueger2012}. The $<$100$>$ oxygen-terminated diamond surface contains primal sp$^2$ configured carbon that forms when dangling bonds relax into carbon-carbon double bonds. There is currently no etching or cleaning method available to remove them permanently as the defects will slowly reform on oxygen terminated surfaces\cite{Stacey2019,Broadway2018}. 
Alternative strategies have been attempted, but are either vulnerable to the ambient environment or show high leakage currents \cite{Davis2017,Abraham2016} making them unsuitable for our applications.\\ 
Since efficient surface passivation is not yet available for diamond, an alternative detector design is proposed here. This design employs s thin conductive layer to extend the on-chip detector electrodes across the entire implantation site (see Fig. \ref{fig:Figure1}). Ions are implanted through the thin conductive electrode (sub-5 nm) with negligible loss of kinetic energy.  In this configuration, the high electric field strength of $10^{7}$ V/m at the implantation site surface repels ion-induced free charge carriers away from the surface before they can become trapped and recombine. Alternative electrode materials suffer from numerous engineering problems \cite{Karakouz2009,Xing2020}. Conductive hydrogen-terminated diamond surfaces are still highly-ohmic, which represents a significant series resistance for the surface electrode that attenuates the electric field strength and thus the charge collection efficiency. Finally a graphene layer does not contribute contamination from forward recoils from non-carbon atoms in the electrode material into the diamond substrate.  \\

In this work, we benchmark graphene electrodes against alternative ultra-thin surface layers fabricated by three different methods. The first device incorporates an oxygen terminated diamond surface which effectively acts as an unpassivated reference.  The second device (referred to as C-implanted), the near-surface carbon lattice was modified to become conductive by means of graphitisation induced by ultra-low-energy carbon implantation ($\approx 2$~keV or below). The third device had a 1 nm thick Pt layer deposited over the implantation site (Pt-deposition layer). The fourth device had a conductive graphene monolayer placed over the implantation site to provide a near-ideal $\sim$ 0.3~nm thin electrode, which is superior in minimising the initial loss and straggling of the kinetic ion energy.\\
The charge collection efficiency of each device was mapped with the MeV IBIC that takes advantage of the deep penetration of MeV ions to allow signals to be induced below the metal on-chip electrodes surrounding the thin electrode area to assess the viability of the fabrications processes to couple the electrodes to the diamond substrate. As will be shown below, the best performing device - equipped with the graphene surface electrode - is then subjected to further kev IBIC tests to evaluate the effectiveness of the graphene layer as well as the single ion detection confidence under conditions required for the shallow colour center formation.
\\
\begin{figure}[t]
    \centering
    \includegraphics[page=2,width=\linewidth,trim={0 1.5cm 0cm 0}]{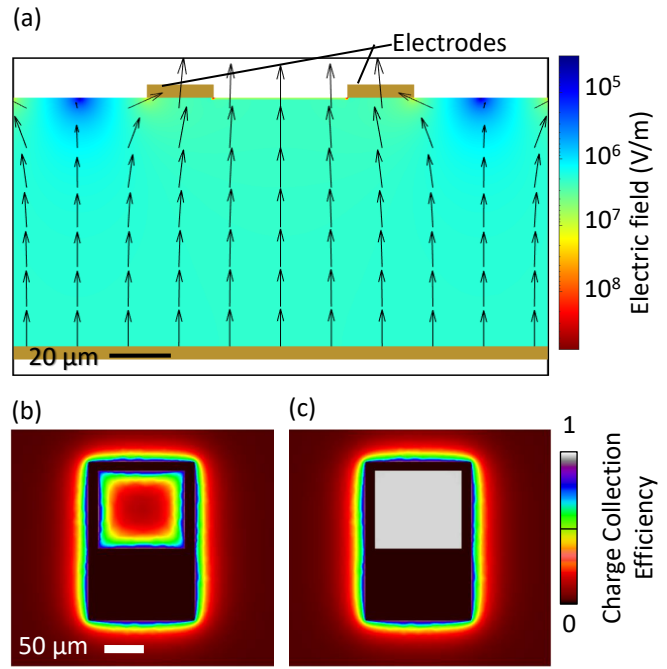}
    \caption{Finite element simulations (COMSOL Multiphysics) of diamond single ion detectors featuring, (a) Cross-sectional electric field distribution for 200~V bias across a 95 \textmu m thick diamond detector substrate. Metal surface electrodes are shown as thick gold stripes and an optional ultra-thin conductive surface layer (e.g. Graphene) connecting them. (b)-(c) Top-view charge collection efficiency maps for 19.5 keV H$^+$ of a diamond detector equipped with (b) with no conductive layer and (c) an ultra-thin ohmic layer inside the implantation site . Device (c) clearly exhibits superior charge collection efficiency and consequently improved single ion detection performance. 
}
    \label{fig:Figure2}
\end{figure}
\section*{Results and Discussion}
\subsection*{Concept and computational predictions}
The electric field distribution inside a sandwich-type diamond detector of the kind illustrated in Fig. \ref{fig:Figure1} and corresponding collection efficiency of free charge carriers that are induced upon 19.5~keV  H$^+$ ion implantation, is numerically computed via a finite element approach (see Experimental Section). Figure \ref{fig:Figure2} illustrates the theoretical predictions for two alternative device configurations - namely, a detector exhibiting an unpassivated implantation site surface and one with an ultra-thin conductive layer covering the implantation site surface. It is evident that the latter design demonstrates a uniform collection efficiency near $\eta=1$  (i.e. 100\%) for free charge carriers that are induced by the low energy ions to a depth of 130~nm of the surface (Fig. \ref{fig:Figure2}c). The ultra-thin conductive surface layer inside the implantation site induces charge drift in the strong electric field that is applied between the conventional metal electrodes and thus suppresses charge carrier diffusion towards the surface where they would recombine. This feature is not present in the other device with a bare surface, which causes a steep decline of charge collection efficiency towards the implantation site centre. 
\begin{figure*}[t]
    \centering
    \includegraphics[page=1,width=\textwidth,trim={0 8.5cm 0cm 0}]{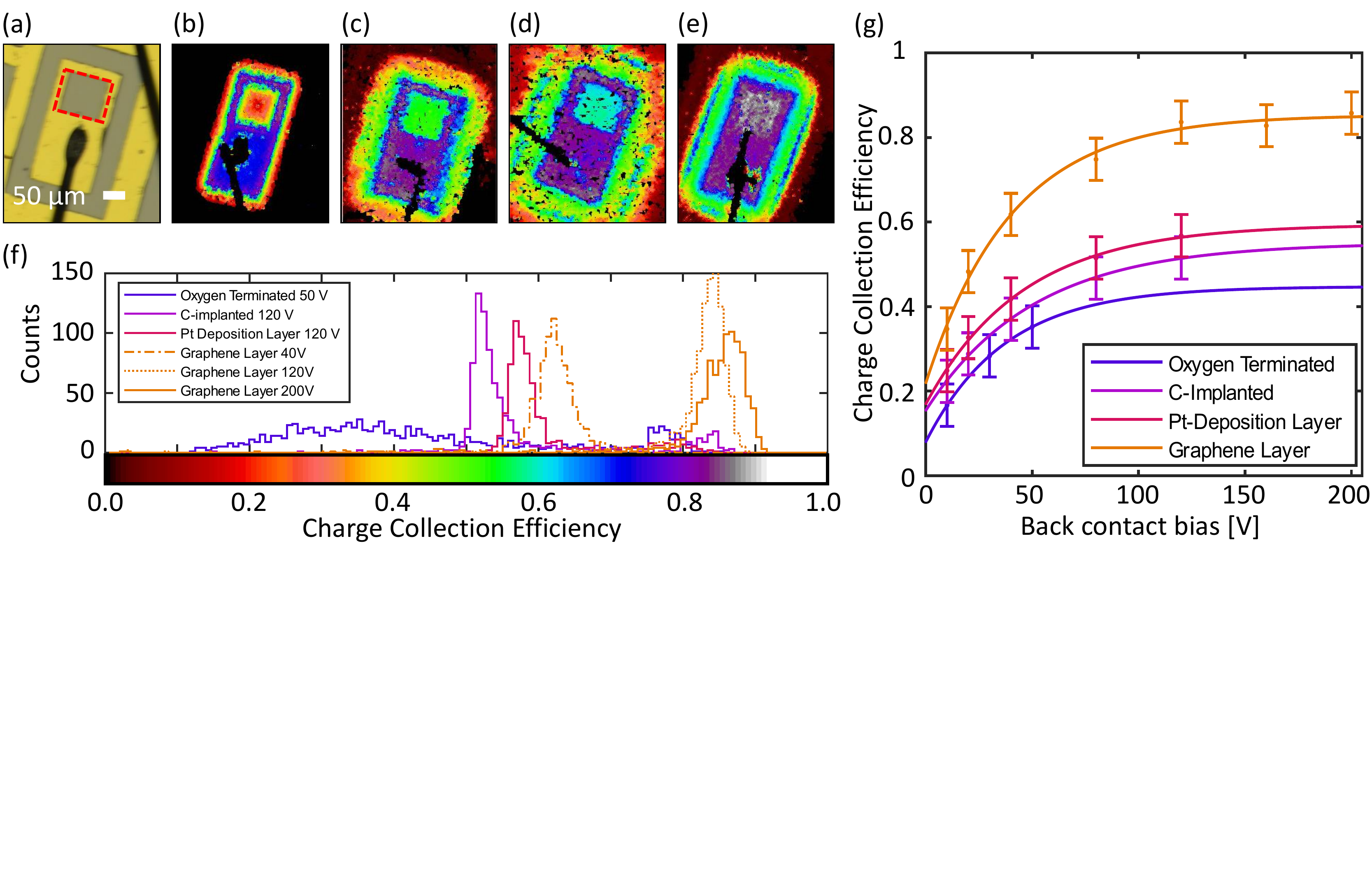}
    \caption{MeV IBIC images of sandwich diamond detectors with different electrodes covering the implantation sites. (a) Optical top-view on detector device 1. (b)-(e) Median IBIC charge signal maps using a scanned focused beam of 1 MeV He$^+$ ion samples 1-4. The implantation site surfaces are (b) O-terminated, (c) low energy C-implanted, (d), thin Pt layer and (e) Graphene layer. The charge pulse maps share a common colour scale which represents the quantitative charge collection efficiency ranging from 0 to 1. (f) The charge pulse height signal spectra of the four detectors. The spectra are extracted from within the implantation sites of the charge signal maps (b)-(e) indicated by the red square in (a). (g) Plot of all MeV charge collection efficiencies with exponential fit for each device as a function of electrode bias voltage showing clearly the highest efficiencies obtained with the graphene device.
}
    \label{fig:Figure3}
\end{figure*}


\subsection*{Performance analysis: Detector electrode materials}
The four detectors of interest were first subjected to spatially resolved charge signal mapping using a scanned focused microbeam of 1 MeV He$^+$ ions from the Melbourne nuclear microprobe system. Here, the charge signal induced by a single ion is very large compared to the detector noise threshold. This approach has the advantage of providing a convenient comparison of the different surface electrode materials even if the efficiency is very low. Figure \ref{fig:Figure3} and table \ref{tab:Table1} summarise the results. The images in Figure \ref{fig:Figure3} map the charge pulse height signal (256$\times$256 pixels) with a microbeam dwell time of 8~ms to keep the total acquisition time acceptable for the given ion current of about 20~ions/s. This timed scanning approach inherently causes some 'dark' pixels, which did not experience an ion impact. This is not a problem for our purposes because we assess the average charge collection efficiency in the implantation site ignoring the dark pixels. 
Device efficiency present in Table \ref{tab:Table1} are extracted via Gaussian curve fitting of the spectra [Fig.\ref{fig:Figure3}f] to find the centre of each peak in the charge pulse height spectrum. This value is then converted into energy using the energy to create an electron-hole in diamond and compared to the expected energy to ionization given by SRIM\cite{Ziegler2010}, with the expected energy lost to the thin conductive layer subtracted from the total energy. For MeV IBIC all samples used detector 1 (Fig.\ref{fig:Figure1}:b) \\
The device with the lowest efficiency (oxygen terminated) in Fig. \ref{fig:Figure1}b, could be only biased to 50~V before electrode breakdown.  The device saturates to its maximum charge collection efficiency of just above 0.4 [Fig.\ref{fig:Figure3}g]. 
Since this device lacks any conductive layer inside the implantation site, the electric field near the surface is very low ( drops exponentially from $\sim$2~V/\textmu m on the metal edge to $\sim$0.03~V/\textmu m in the center of the O-terminated implantation site at 50 V). As a result, the implantation site in Fig. \ref{fig:Figure3}b) features a pronounced lateral deterioration of the charge collection efficiency towards its center with increasing distance from the surrounding metal electrode. This experimental result is in good agreement with the computed charge collection efficiency map shown in Fig. \ref{fig:Figure2}b) for an analogously configured implantation site (no conductive surface electrode). However, to achieve this quantitative agreement, the simulation assumes a charge carrier bulk lifetime of only $5\times10^{-8}$~s, which is not realistic for electronic grade diamond substrates ($1\times10^{-6}$~s) and points towards surface recombination effects. Accordingly, the experimental charge collection efficiency spectrum of the O-terminated device (see blue plot in Fig. \ref{fig:Figure3}f) is characterised by a broad shallow peak extending from about 0.1 to 0.6 (or correspondingly efficiencies 10\% to 60\%). 
\begin{table}[b]
\begin{tabular}{p{0.9cm}p{1.7cm}p{1.7cm}p{1.7cm}p{0.5cm}p{0.5cm}}
\hline
Sample & Electronic Material & Electrode Thickness {[}nm{]} & 1 MeV H$^+$ Charge Loss {[}\%{]} & CCE {[}\%{]} & Bias {[}V{]} \\ \hline
O                  & Oxygen terminated                & 0                        & 0                             & 30.5                           & 50           \\
C                  & C-implanted                      & 4                        & 0.27                          & 52.2                           & 120          \\
Pt                 & Pt-layered                       & 1.0                      & 0.063                         & 57.7                           & 120          \\
G                  & Graphene                         & 0.4                      & 0.023                         & 61.8                           & 40           \\
G                  & Graphene                         & 0.4                      & 0.023                         & 83.6                           & 120          \\
G                  & Graphene                         & 0.4                      & 0.023                         & 85.9                           & 200          \\ \hline
\end{tabular}
\caption{Summary of charge signal maps using a scanned 1 MeV He$^+$ ion beam. The data refers to device characteristics inside the implantation sites. }
    \label{tab:Table1}
\end{table}
The C-implanted and Pt-layer devices showed relatively uniform charge collection efficiency at 120~V electrode bias (see Figs. \ref{fig:Figure3}c and d). This result can be also extracted from the two charge collection efficiency spectra shown in Fig. \ref{fig:Figure3}f. Both spectra contain practically all charge signal events in a sharp confined peak located at 52\%  and 58\% , respectively. A laterally homogeneous detector signal response is generally required to maintain a high single ion detection confidence and allowing charge signal spectrometry employed for estimating the implantation depth of single ions from the known ion-solid interaction physics. The slightly improved charge collection efficiency of the Pt layer device is attributed to a lower conductivity of the graphitic diamond surface compared to device 4 ($\sim 9.5\times10^6$~S/m for bulk Pt vs $\sim 2\times10^2$~S/m for graphite \cite{Zhang2020,falcao2007}) and an inferior ohmic contact with the surrounding conventional metal top electrode. As a result, the electric field strength between the C-implanted implantation site surface and the back electrode is slightly more attenuated, causing a higher charge carrier fraction to recombine near the surface. Alternative methods that may improve results for these samples include B doping to achieve a higher conductivity with a lower implant energy and higher dose. Though this would require alternative fabrication process 
Finally, the graphene layer device, allowed a relatively  high electrode bias operation point of 200~V, resulting in notable 2~V/\textmu m field strength (7~V/\textmu m under the graphene) between top and back electrodes and the highest average of 86\% charge collection efficiency inside its implantation site. The fact that this device did not converge towards 100\% with even higher electrode bias suggests that its internal sheet resistance is not low enough for a true low-ohmic behaviour. Alternatively the electrical contact with the surrounding metal electrode or the diamond below is imperfect with residual series resistance at the interface\cite{Bogdanowicz2019}. Indications supporting the former idea can be taken from its charge collection efficiency spectrum (blue plot in Fig. \ref{fig:Figure3}f), which exhibits a relatively broad peak compared to the C-implanted and Pt-layer devices. Accordingly, corresponding map [Fig. \ref{fig:Figure3}e] shows a speckle-like fluctuation of local charge collection efficiency rather than being completely homogeneous. This could be either attributed to nano-scale debris left from fabrication or structural inhomogeneity of the graphene layer itself. The likely origin will be discussed further below by means of scanned keV IBIC that is able to probe the cause of these fluctuations with improved sensitivity. 
\\
\subsection*{Graphene Electrode Device: Detector performance for keV ions}
\subsubsection*{Qualitative High-Resolution FIB mapping}
\begin{figure}[t]
    \centering
    \includegraphics[page=3,width=\linewidth,trim={0 0.5cm 0cm 0}]{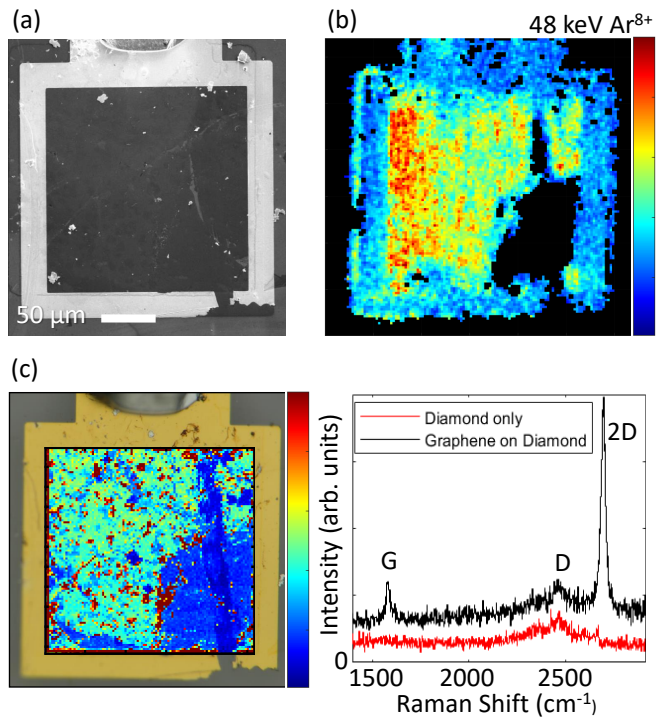}
    \caption{200$\times$200 \textmu m$^2$ size implantation site. a) SEM of the detector. b) 8$^+$ 6 keV Ar IBIC median energy map. c)  Raman mapping of the graphene 2D peak overlaid over optical image of graphene detector 2 [Fig. \ref{fig:Figure1}]. Insert shows sample spectra from different locations with G, 2D and 2$^{\mathrm{nd}}$ order Raman lines for device locations with and without graphene. Each pixel corresponds to a 2 by 2 \textmu m$^2$ area.  
}
    \label{fig:Figure 4}
\end{figure}
Focusing on the best performing device with the graphene layer, spatially resolved charge collection efficiency mapping was conducted under  conditions similar to those required for the engineering of shallow implanted dopant networks for quantum devices, i.e. with keV ions. A 48 keV Ar$^{8+}$ focused ion beam (FIB) of 0.8 \textmu m spot size was scanned across the detector implantation site. Similar to the MeV ion charge collection mapping, a timed pixel stepping approach was chosen with a pixel dwell duration of 8 ms and $\sim$80 ions/s beam current. The average ion implantation depth for this species and energy amounts to about 28 nm in diamond, proving charge signal information from free carriers induced in near vicinity beneath the graphene surface electrode. 
\\
Figure \ref{fig:Figure 4} summarises results for a selected graphene electrode device that exhibits an implantion site area of 200$\times$200 \textmu m$^2$ (Fig.\ref{fig:Figure1}:b device 2) which is significantly larger than those presented in the previous section and therefore much more sensitive to variations of the internal electric field induced by the surface electrode. This can be seen by comparing the scanning electron microscope image  [Fig.\ref{fig:Figure 4}a] of the device with the 48 keV Ar$^{8+}$ IBIC map [Fig.\ref{fig:Figure 4}b] which shows variations in the charge collection efficiency across the device with a gradient from left to right. This suggests either that the electrical contact between graphene and gold electrode or diamond substrate is non-uniform or there are spatial variations within the graphene electrode itself such as variations in the sheet resistance arising from local distortions of the bi-axial strain, or sheet cracks and the interface between graphene and diamond surface. A high sheet resistance results in attenuation and distortion of the external drift field, which in turn promotes charge carrier trapping and reduced charge collection efficiency. Furthermore, a large patch is dark with no IBIC signal (i.e. signals below the noise threshold discriminator ) can be observed close to the bottom right corner. Raman spectrometry mapping, shown in Fig. \ref{fig:Figure 4}:cshows the graphene peak at 2680 cm$^{-1}$ Raman shift is absent from the dark area showing the graphene electrode is missing from this region. The Raman map also reveals small patches at the edge of the dark region with an increased 2D peak (dark red pixels). This suggest the graphene layer is thicker at the edges of the dark area as a consequence of device processing trauma that caused the graphene electrode to be dislodged from the surface and partially peeled back at the edges of the darks region.  Although this is undesirable for a fully functioning device, it nevertheless shows the clear role of the intact graphene electrode in producing the high charge collection efficiency observed in the undamaged regions. 
\\ 
One further important observation can be made from the IBIC map of this device which is the presence of IBIC signals from the thick gold electrode that frames the graphene electrode [Fig. \ref{fig:Figure 4}b]. Given that it is not possible for the 48 keV Ar$^+$ ions to penetrate through the 130 nm thick gold electrode to induce free charge carriers inside the diamond bulk, these signals do not arise from the IBIC process.  Instead these signals arise from the induction of secondary electrons from the surface of the device from the highly charged Ar$^{8+}$ ions.  This process results in a current transient that is above the noise discriminator threshold of the pre-amplifier.  This an attribute of the high sensitivity of the pulse processing electronics and may have future applications to the investigation of electron emission from the metalisation of small-scale devices.  
\\

\subsubsection*{Quantitative charge signal analysis}

\begin{figure*}[t]
    \centering
    \includegraphics[page=2,width=\textwidth,trim={0 0cm 0cm 0}]{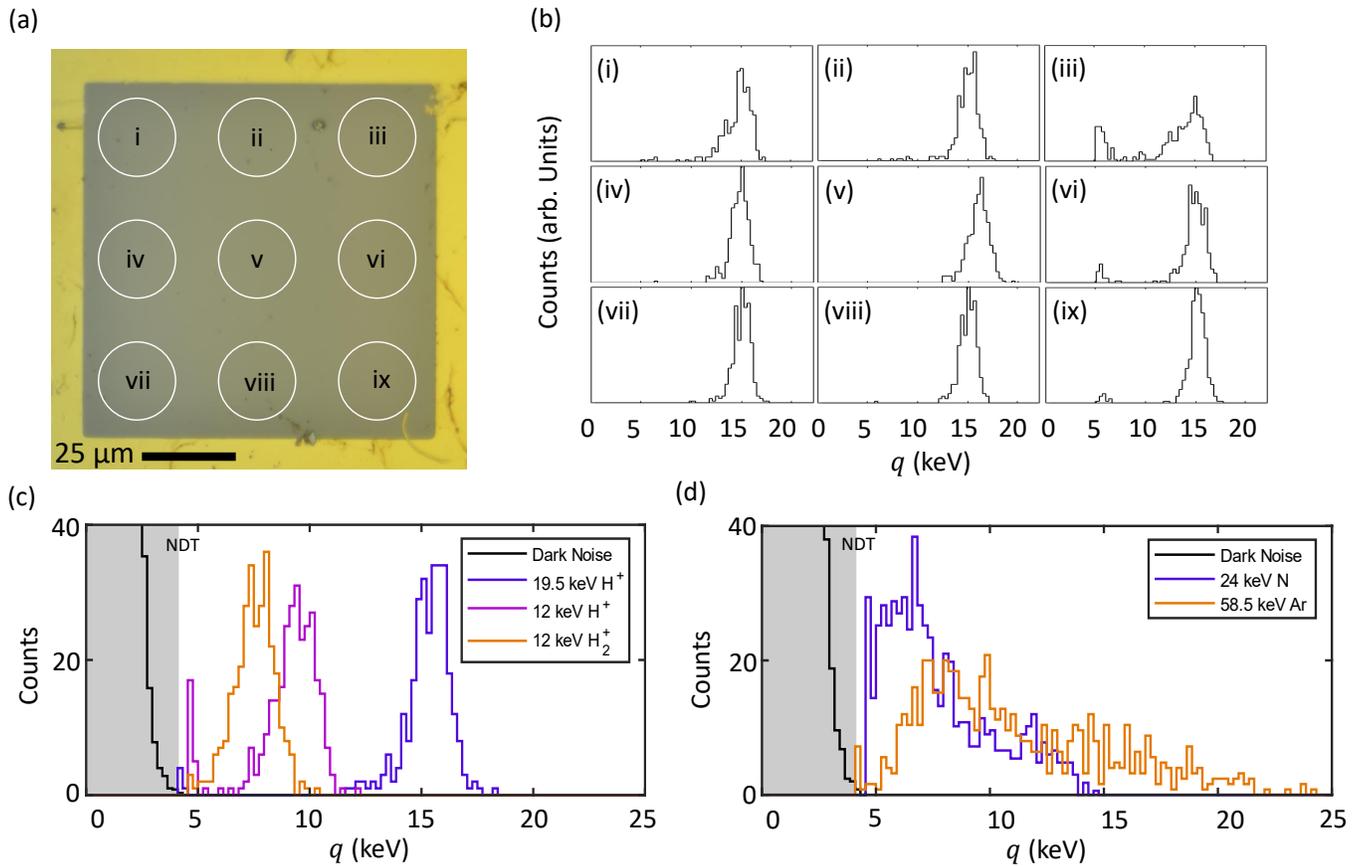}
    \caption{IBIC spectra employing keV ions from a diamond detector with a graphene front electrode biased to 165 V: (a) Optical micrograph of detector prior to IBIC showing nine 22 \textmu m diameter regions of interest; (b) corresponding IBIC spectra from the regions of interest obtained with 300 19.5 keV H$^+$ ions; (c) light ion IBIC spectra (19.5 keV H$^+$, 12 keV H$_2^+$ and H$^+$) showing relatively high CCE; (d) heavy ion IBIC spectra ( 24 keV N$^+$ and 58.5 keV Ar$^+$). Both (c) and (d) where done over region (v) and also show dark noise spectra of no beam measured with lower noise discriminator threshold to show the signals excluded from the IBIC spectra by the lower level discriminator in the data acquisition system. 
}
    \label{fig:Figure5}
\end{figure*}
IBIC signals from a device with an intact graphene electrode are shown in [Fig. \ref{fig:Figure1}:b Implantation site 3] employing a series of light and heavy ion species to investigate the lateral uniformity and the efficiency as a function of ion mass. By means of H$^+$ and H$_2^+$ ions with implantation energies ranging from 6 keV to 19.5 keV a depth profile of the detector's charge collection efficiency can be extracted. Also, to investigate the lateral variations, the 100$\times$100 \textmu m$^2$ implantation site was targeted with a 3$\times$3 grid [Fig. \ref{fig:Figure5}a]  of nominally 22 micon diameter regions of interest (ROI) that were individually irradiated with 300 subsequent single 19.5 keV H$^+$ ions.   The IBIC spectra from each ROI allowed comparison of the uniformity of the device. The results are shown in Fig. \ref{fig:Figure5}b and reveal consistent charge collection efficiency for 7 out of the 9 ROIs within the statistical variation. The top-left and top-right segments exhibit broader peak widths and lower charge signal amplitudes, which is consistent with energy loss in adventitious surface debris or unpassivated surface charge traps. Noteworthy is the fact that the central ROI shows an about 5\% above average charge collection efficiency of $\eta=0.8$ (as extracted from the peak center position) compared to all other segments. Once again, this shows that the surface graphene electrode maintains a strong electric field across the whole area of the device.     
\begin{table}[b]
    \centering
    \begin{tabular}{p{1.5cm}p{1.5cm}p{1.5cm}p{1.5cm}p{0.9cm}}
\hline
Ion Species            & Energy [keV] & Range [nm] & Mean N$_{e-h}$ & CCE [\%]   \\ \hline
H$^+$                  & 19.5   & 130.7      & 1400           & 82.1 \\
H$^+$ & 12     & 90.6       & 900            & 81.3 \\
H$_2^+$                & 12     & 53.7       & 850            & 69.1 \\
N$^+$                  & 24     & 32.4       & 1250           & 44.6 \\
Ar$^+$                 & 58.5   & 33.1       & 2800           & 23.0 \\ \hline
\end{tabular}
    \caption{Summary of the IBIC measurements from the 100x100 \textmu m device with  12 keV N$^{2+}$ and 19.5 keV Ar$^{3+}$ ions.  The N$^+$ and Ar$^+$ CCE values reach up to 70 \% due to ion channelling. 
    }
    \label{tab:Table2}
\end{table}
\\
\\
This central ROI was used for further quantitative charge signal experiments with a series of H$^+$ ions and H$_2^{+}$ molecule-ions at only 12 keV kinetic energy.  The average ion implantation depth was effectively reduced to 90 nm (H$^+$) and 53 nm (H$_2^+$) compared to the initial irradiation's with 19.5 keV H$^+$ ions (130 nm). Consequently, the free charge carrier induction from the ion electronic stopping fraction occurs closer to the surface. As the implantation energies of 12 keV H$^+$ ions and 12 keV H$_2^+$ molecule-ion ($\cong2\times$6keV H$^+$) is the same, it would be expected that the IBIC spectra of both species to be similar, apart from minor differences in the stopping powers. However, as shown in Fig. \ref{fig:Figure5}d, the IBIC spectrum of the much shallower implanted H$_2^+$ molecule ion species shows a clear shift towards lower signal amplitudes. Whilst the charge collection efficiency of the device under test reduces only from 0.82 for 19.5 keV H$^+$ ions to 0.81 for 12 keV H$^+$, the efficiency of 0.69 for molecules with 6 keV per atom becomes more apparent. This non-linear behaviour shows the proximity to the surface results in greater charge carrier recombination. This is a consequence of the radially isotropic charge carrier diffusion towards the surface is not compensated by the uni-directional carrier drift away from the surface from the bias field applied to the graphene surface electrode. Additional effects like a fixed surface charge, due to trapped interface defects, cannot be excluded and would attenuate the near-surface drift field further. 
Hence, improving the collection efficiency for free charges carriers induced sub-50 nm deep will require further modification of the device fabrication process. For a further advanced next-generation detector approaching 100\% charge collection efficiency, a two- or multi-layer graphene electrodes that may be additionally chemically doped would reduces the electrode sheet resistance \cite{Bianco2020} and improve the homogeneity and strength of the bias field.  IBIC spectra obtained with heavy ions are shown in Fig. \ref{fig:Figure5}d, obtained from the central ROI with 1000 24 keV N$^{2+}$ ions and 750 58.5 keV Ar$^{3+}$ ions. Both spectra peaks are characterised by a pronounced tail towards higher charge signal values (up to 15 keV for N and 25 keV for Ar). This is due to the incident ion beam being aligned with the <001> axis of the diamond substrate. As a consequence, ion channeling occurs, where a significant fraction of ions are captures into channels in the crystal lattice. The kinetic energy is then predominantly dissipated via electronic stopping rather than nuclear collisions. These ions are implanted much deeper compared to ions subject to random collisions in the diamond lattice, which is undesirable for consistent shallow dopant network engineering. The excess electronic stopping fraction of such channeled ion implantation events results in an increased induction of free charge carriers and causes a much larger charge signal (as evident by obtained spectra tail). 
This effect can be reduced by introducing a small tilt correction (i.e. $\approx 7^\circ$) between detector and the incident ion beam to suppress the channeled fraction of ions. It is also evident that, for the present device, a significant portion of the 24 keV N$^{2+}$ IBIC spectrum overlaps with the dark noise spectrum. The IBIC spectrum below the lower level discriminator of 4.5 keV is discarded and limits the single ion detection confidence. Quantifying this fraction requires knowledge of the entire signal spectrum shape to precisely determine the convoluted area with the noise spectrum. As this detector does not exhibit 100\% charge collection efficiency, a reliable modelling of the signal spectrum from first principle simulations \cite{Pilz1998, Jakob2022} is not feasible. However, using a Gaussian fit of the IBIC spectrum allows a first order estimate of the single ion detection confidence (see the Experimental Section) to approximately 85.5$\pm$2.5\%. In contrast, the signal spectrum from 750 single 58.5 keV Ar$^{3+}$ ion events is clearly separated from the dark noise spectrum. The occurrence of charge signal events detected below 5 keV is - within the statistical error - as abundant as the dark noise event spectrum, where the ion beam is blanked off (see Fig. \ref{fig:Figure5}d)  
An analogous first-order Gaussian fit results in an estimate single ion detection confidence of nominally 98.2$\pm$0.3\%. 
This finding demonstrates that the present device is already suitable for the deterministic engineering of small networks of near surface ($\approx$20-40 nm) dopant species up to 40 amu (e.g. $^{13}$C, $^{14}$N, $^{28}$Si).

\section*{Conclusions}
On the basis of theoretical modelling of free charge subject to the electric field within a diamond sandwich detector we determined that, without a surface electrode at the ion impact site, charge recombination degrades the IBIC signal needed for deterministic ion implantation. However we demonstrated an atomically thin graphene surface electrode addressed this problem by drifting the free charge away from the surface traps. The atomically thin layer provides minimal incident ion kinetic energy loss whilst the high conductivity ensures the bias field in propagated into what would otherwise be field-free regions.  In combination with a low-noise charge sensitive pre-amplifier this allows deterministic implantation of sub 50~keV ions needed to investigate the fabrication of near-surface diamond colour centres. For light ions (H) down to 6~keV, signals can be detected above the noise threshold with high confidence (99.9\%). For 24~keV N ions, the detection confidence was 85.5\% which is sufficient for the first step of addressing the challenge of fabricating colour centre arrays. This method can be adapted for placing single group IV  or other ions into diamond at the suitable energies which can then be post-processed into waveguides or resonators\cite{Nguyen2019,Nemoto2016,Bradac2019}. 
\\
Further improvements can be made to reduce the leakage current, as well as to achieve higher charge collection efficiency. Hydrogen terminated diamond also remains as a potential thin conductive source, that would have no dead layer and can be fabricated with minimal defects.


\section*{Experimental Section}
\subsubsection*{Sample Preparation}
Diamond substrates with dimensions 4~mm $\times$ 4~mm $\times$ 0.1~mm were sourced from Delaware Diamond Knives and cut into quarters using a Oxford alpha series laser cutter. The samples were subsequently cleaned using an acid boil solution (sulphuric acid and sodium nitrate) for 60 minutes. A sequence of Reactive Ion Etching (RIE) (Ar/Cl at 5/30 sccm 100/1000 w RF/ICP power for 2~hrs, followed by 2 minutes of O$_2$) was employed to etch 2~\textmu m of diamond surface material for further surface cleaning. A stacked 20/10/100~nm Ti/Pt/Au layer was deposited via electron beam evaporation on to create the uniform back electrode. The front electrode pattern was fabricated with photolithography using the same metalisation scheme.\\
Device 1 was not processed any further, leaving it with an oxygen-terminated front surface for ion implantation surrounded by the metalisation. 
Device 2 was masked via photolithography, leaving only the ion implantation site exposed.  This was then irradiated with 10 keV $^{12}C_8^{-}$ molecule ions (${10^{15}}$~cm${^{-2}}$ fluence) at the Australian National University to induce a surface damage layer from the shallow implantation depth of 2.95~nm according to TRIM simulations \cite{Ziegler2010}. A subsequent anneal at 800$^\circ$C for 2~hrs in vacuum, followed by a 20~min oxygen plasma clean forms a $\sim$4~nm graphitic layer at the diamond surface over the implantation site. 
Device 3 was likewise masked via photolithography to expose only the implantation site and subsequently sputter-coated with a 1~nm Pt layer.
Device 4 received a graphene surface layer deposited with a wet transfer method. Subsequently the ion implantation site was protected with a photolithographic mask and then subject to an oxygen plasma clean for 20~mins to remove the graphene layer in all regions, except the implantation site and the edges of the surrounding metalisation. \cite{Chen2016}

\subsubsection*{IBIC Signal Analysis}
The Melbourne nuclear microprobe was used for IBIC with 1 MeV He$^+$ ions. The focused ion beam ($\sim$1~\textmu m) was scanned across the devices and the median energu of the IBIC signal at each pixel was used to make the map of charge collection efficiency. For IBIC measurements with keV ions, the Melbourne Colutron ion implanter was employed with ions directed to regions of interest in the devices by means of a AFM stencil.  IBIC spectra were obtained using 19.5 keV H$^+$, H$_2^+$, He$^+$, Ar$^{3+}$ and N$^+$ ions \cite{Jakob2022}. The beams were collimated to 22 \textmu m diameter by the stencil. Additional IBIC spectra were obtained by the Leipzig IOM Focused Ion Beam system\cite{Racke2020} to implant 6~kV Ar$^{2+}$, Ar$^{4+}$, Ar$^{6+}$ and Ar$^{8+}$ focused ion beams with 1~\textmu m diameter beam scanned across the implantation sites to produce IBIC median energy maps \cite{Robson2022}. 
IBIC spectra from the regions of interest were obtained from the signals tagged with the beam location on the device by the data acquisition system.  Gaussian curve fitting the IBIC signal peak was used to calibrate the charge collection efficiency with reference to known high efficiency silicon detectors allowing for the energy needed to create a single electron-hole pair arising from the different band gaps of silicon and diamond. The IBIC spectra typically showed high energy tails from a fraction of ions captured into crystal channels and in this case two Gaussian curves were employed to fit the random and channeled populations.  Only the peak from the random population of iosn was used for the calibration. 
\subsubsection*{Surface Characterisation}
Raman spectroscopy was conducted on the graphene electrode of device 4 using a Renishaw inVia Qontor confocal Raman microscope. A laser of 532~nm wavelength and $\sim$1~\textmu m spot size was scanned over the implantation site with a 200$\times$200~\textmu $m^2$ scan in 2 \textmu m steps. A Raman spectrum was obtained from each pixel in the scan. The Raman shift was scanned from 1400 to 2900 cm$^{-1}$ to map the graphene signal. The Raman spectra were calibrated from the diamond Raman line at 1332 cm$^{-1}$

\subsubsection*{Simulated Electronic Device Properties}

COMSOL Multiphysics was employed to simulate the electric field distribution inside a diamond detector as well as the collection efficiency of free charge carriers induced by near surface implanted ions. The collection efficiency was computed using the well-established weighting field approach introduced in \cite{Vittone2008}. The Comsol Multiphysics semiconductor module employs Gunn's theorem \cite{Gunn1964} to generate the weighting field and local charge carrier lifetimes. The spatial charge collection efficiency is then derived using the electronic stopping distribution for the ion species of interest and extracted from TRIM \cite{Ziegler2010}. The initial loss of the kinetic ion energy when passing through metal electrode layers was included in the simulation.

\medskip

\medskip
\textbf{Acknowledgements} \par 
This work was funded by the Australian Research Council center of Excellence for Quantum Computation and Communication Technology (Grant No. CE170100012), the U.S. Army Research Office (Contract No. W911NF-17-1-0200) and the International Atomic Energy Agency Cooperative Research Program, Ion beam induced spatio-temporal structural evolution of materials: Accelerators for a new technology era, CRP No: F11020. This work was performed in part at the Melbourne center for Nanofabrication (MCN) in the Victorian Node of the Australian National Fabrication Facility (ANFF). We acknowledge access to the NCRIS facilities (ANFF and Heavy Ion Accelerator Capability) at the Australian National University. N.F.L.C acknowledges additional travel support from the Laby Foundation. N.F.L.C. and S.G.R. acknowledge support from an Australian Government Research Training Program Scholarship. P.R. and D.S. gratefully acknowledge funding by the Leibniz Association (SAW2015-IOM-1) and the European Union, together with the Sächsisches Ministerium für Wissenschaft und Kunst (Project No. 100308873). PR gratefully acknowledges funding by the Federal Ministry of Education and Research under grant no. 13N16097 (CoGeQ).

\medskip
\textbf{Conflict of Interest} \par
The authors declare no conflict of interest.

\medskip
\textbf{Author Contributions} \par

\medskip
\textbf{Data Availability Statement} \par
The data that support the findings of this study are available from the corresponding author upon reasonable request.

\medskip

\bibliography{mendelay.bib}
\end{document}